\documentclass[twocolumn, journal]{IEEEtran}
\usepackage{color}
\usepackage{graphicx}
\usepackage{epstopdf}
\usepackage{amsmath}
\usepackage{amssymb}
\usepackage{algorithm}
\usepackage{algorithmic}
\usepackage{amsmath}
\usepackage{multirow}
\usepackage{booktabs}
\usepackage{array}
\usepackage{amsthm}
\usepackage{stfloats}
\usepackage{caption}
\usepackage{subfigure}
\usepackage{bm}
\usepackage{booktabs}
\usepackage{setspace}
\usepackage{gensymb}
\usepackage{url}

\newcommand{\be}{\begin{equation}}
\newcommand{\ee}{\end{equation}}
\newcommand{\bea}{\begin{eqnarray}}
\newcommand{\eea}{\end{eqnarray}}
\newcommand{\ba}{\begin{array}}
\newcommand{\ea}{\end{array}}

\allowdisplaybreaks[4]

\title{Reflection and Relay Dual-Functional RIS Assisted MU-MISO Systems
\thanks{This work is supported in part by the National Natural Science Foundation of China (Grant No. 61971088, 62071083, U1808206, and U1908214), the Natural Science Foundation of Liaoning Province (Grant No. 2020-MS-108), in part by the Fundamental Research Funds for the Central Universities (Grant No. DUT20GJ214, DUT21GJ208 and DUT20RC(3)029), in part by Dalian Science and Technology Innovation Project (Grant No. 2020JJ25CY001),  and  in part by the Open Research Fund of National Mobile Communications Research Laboratory, Southeast University (Grant No. 2021D08). }}
\vspace{-0.0 cm}
\author{\IEEEauthorblockN{Yanan Ma$^*$, Rang Liu$^*$, Ming Li$^{*\dag}$, Yang Liu$^*$, Qingqing Wu$^{\ddag}$ and Qian Liu$^*$}\\
\IEEEauthorblockA{$^*$Dalian University of Technology, Dalian, Liaoning 116024, China \\ E-mail: \texttt{\{mayanan, liurang\}@mail.dlut.edu.cn, \\
\{mli, yangliu\_613, qianliu\}@dlut.edu.cn} }\\
\IEEEauthorblockA{$^\dag$Southeast University, Nanjing, Jiangsu 210096, China}\\
\IEEEauthorblockA{$^{\ddag}$University of Macau, Macau, 999078, China \\ E-mail: \texttt{qingqingwu@um.edu.mo} }
}
\begin{document}

\maketitle

\pagestyle{empty}
\thispagestyle{empty}

\begin{abstract}
Reconfigurable intelligent surface (RIS) is a promising solution to adaptively manipulate wireless propagation  with low-cost passive devices. However, the traditional passive RIS can offer sufficient signal strength only when receivers are very close to it. Moreover, the users at the back side of it cannot be well served due to its reflective property. In this paper we introduce a novel reflection and relay dual-functional RIS architecture, which can simultaneously realize passive reflection and active relay functionalities. The problem of joint transmit beamforming and dual-functional RIS design is investigated to maximize the achievable sum-rate of a multiuser multiple-input single-output (MU-MISO) system.
Based on fractional programming (FP) theory and majorization-minimization (MM) technique, we propose an efficient iterative transmit beamforming and RIS design algorithm.
Simulation results demonstrate the superiority of the introduced dual-functional RIS architecture and the effectiveness of the proposed algorithm.
\end{abstract}

%
%\maketitle
\vspace{-0.0 cm}
\section{Introduction}
%\vspace{-0.1 cm}
Reconfigurable intelligent surface (RIS) has been proposed as an inherent component of future wireless communication systems owing to its superior ability of shaping the wireless environment \cite{Wu}-\cite{Hybrid}. However, due to ``double fading'' effect (i.e., the equivalent path loss of the transmitter-RIS-receiver link is the product of the path losses of transmitter-RIS link and RIS-receiver link, which is usually an order-of-magnitude larger than that of the direct link), RIS should be deployed close to either the base station (BS) or users to effectively improve communication quality-of-service (QoS) \cite{Wu}, \cite{Active}. Besides, the research on the comparison of RIS and relay illustrates that RIS can be easily outperformed by conventional full-duplex (FD) amplify-and-forward (AF) relay unless very large RIS is employed \cite{Bjornson}, \cite{Ntontin}. However, very large number of reflecting elements will lead to high training overhead for channel estimation and high complexity of beamforming designs \cite{You}. Meanwhile, the
control of a large RIS will also become a difficult task.
%Meanwhile, the control of RIS will also become more difficult.

Moreover, the users at the back of RIS cannot be well served due to its reflective property. The service coverage of RIS is limited to only one side of the surface. To overcome this limitation, simultaneously transmitting and reflecting RIS (STAR-RIS) was proposed \cite{STAR}, \cite{STAR2}, which can reflect and transmit the incident signals to serve the users in front of RIS or behind it.
%More specifically, STAR-RIS can reflect part of the incident signals to the users at the same side, while allowing part of the incident signals to penetrate to the users at the other side.
Thus, the STAR-RIS can provide service for users at both sides of it and improve coverage in a reflective and transmissive way \cite{wenhao}. However, the performance improvement of passive STAR-RIS is still limited due to severe signal attenuation.

In order to overcome the double fading and extend the coverage, the authors in \cite{relay} proposed a novel relay-aided RIS architecture consisting of two RIS surfaces connected via an FD relay, which can achieve the same performance with conventional RIS while only requiring much fewer reconfigurable elements. However, they only consider single-input single-output (SISO) system and derive the theoretical upper bound of achievable rate. The designs of BS beamforming and RIS reflection have not been well investigated.
Another limitation of this work is that the first RIS is adjusted to reflect the signal towards the relay only and cannot simultaneously serve the users around it, which does not take the full use of RIS.
In \cite{Hybrid}, \cite{Hybrid2}, the authors proposed a semi-active RIS-aided architecture, which can not only passively reflect the signal, but also actively amplify it at the same time. However, the practical implementation of this hybrid-RIS remains an open problem, while it needs expensive radio frequency (RF) chains. Active RIS \cite{Active}, \cite{liang} has recently been proposed to overcome the practical implementation problems of passive RIS. However, the hardware implementation is still an issue needs to be addressed.
%
%
%%Relay can increase the energy of received signals, nevertheless, it will increases the interference and can not focus the signal to intended receiver. For the purpose of combining these benefits and overcoming these challenges,

In this paper we propose a novel reflection and relay dual-functional RIS architecture to overcome the double fading effect, enhance the coverage and reduce hardware complexity. The dual-functional RIS unites two RISs back-to-back with a full-duplex amplifier and can simultaneously realize passive reflection and active relay functionalities. Particularly, one RIS passively adjusts its phase shifters to serve the users close to it and at the same side as this RIS. Meanwhile, the signal is amplified and transmitted by the other RIS to serve the users far away from BS/RIS and at the back side of it. Based on the proposed dual-functional RIS architecture, a sum-rate maximization problem is considered for an MU-MISO system. A joint transmit beamforming and RIS reflection design algorithm is developed based on fractional programming (FP) theory and majorization-minimization (MM) technique.
%The convergence of the proposed algorithm is illustrated by simulation results.
The superiority of the dual-functional RIS architecture and the effectiveness of the proposed design algorithm are validated by simulation results.

\vspace{-0.0cm}
\section{System Model}
%\vspace{0.2 cm}

\begin{figure}[t]
\centering
\includegraphics[height=2.1 in]{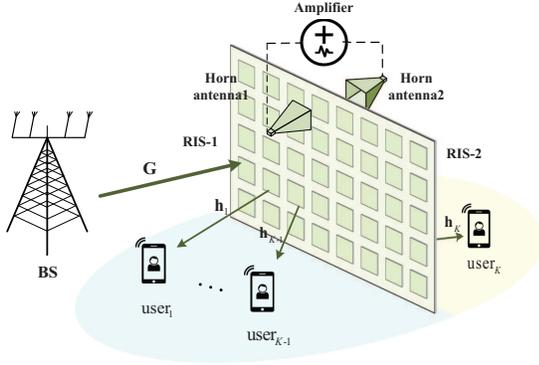}
\caption{The proposed reflection and relay dual-functional RIS architecture in an MU-MISO system.}\label{fig:system}
\vspace{-0.25 cm}
\end{figure}

We consider dual-functional RIS assisted wireless communication system as illustrated in Fig. \ref{fig:system}. The BS equipped with $N$ antennas communicates to $K$ single-antenna users with the aid of a dual-functional RIS. Specifically, the proposed dual-functional RIS is composed of two conventional passive RISs connected via a full-duplex amplifier. Each conventional RIS comprises a large number of passive elements, each being able to reflect the incident signal with a controllable phase-shift. As shown in Fig. \ref{fig:system}, user-1, \ldots, user-${(K-1)}$, are located around RIS and served by the RIS facing them (which is labelled as RIS-1 in Fig. \ref{fig:system}).
Meanwhile, the $K$-th user cannot receive sufficient strong signals from the BS or RIS-1 due to the fact that it is located in the opposite of it and the direct link to BS is blocked. Therefore, we propose the dual-functional RIS scheme to provide satisfactory QoS to all users, no matter where they are located. Specifically, the signal from BS is first reflected by RIS-1 to horn antenna-1, then amplified and emitted to the second (right) side of the RIS (labeled as RIS-2), which transmits the signal towards the $K$-th user. We assume that there exists no direct link from RIS-1 to the $K$-th user and vice versa.

\begin{figure*}[!t]
\normalsize
\setcounter{equation}{9}
\begin{equation*} \label{eq:reR1}
g_\mathrm{R}(\boldsymbol{\Phi}_{1},\boldsymbol{\Phi}_{2},\mathbf{w}_{k},\bm{\gamma},\bm{\tau}) =\sum_{k=1}^{K} \log _{2}\left(1+\gamma_{k}\right)-\sum_{k=1}^{K}\gamma_{k}\\
+\sum_{k=1}^{K-1}(2\sqrt{1+\gamma_{k}}\Re\left\{\tau_k^{\ast}\mathbf{h}_{k}^{H} \boldsymbol{\Phi}_{1} \mathbf{G}\mathbf{w}_{k}\right\}-\left|\tau_k\right|^2\sum_{i=1}^{K}\left|\mathbf{h}_{k}^{H} \boldsymbol{\Phi}_{1} \mathbf{G}\mathbf{w}_{i} \right|^{2}-\left|\tau_k\right|^2{\sigma_{k}^{2}})\\
\end{equation*}
\begin{equation} \label{eq:reR2}
+2\sqrt{\beta\left(1+\gamma_{K}\right)}\Re\left\{\tau_K^{\ast}\mathbf{h}_{K}^H\boldsymbol{\Phi}_{2} \mathbf{g}_\mathrm{r} \mathbf{g}_\mathrm{t}^{H} \boldsymbol{\Phi}_{1} \mathbf{G}\mathbf{w}_{K}\right\}-\left|\tau_K\right|^2\beta\sum_{i=1}^{K}\left|\mathbf{h}_{K}^H\boldsymbol{\Phi}_{2} \mathbf{g}_\mathrm{r} \mathbf{g}_\mathrm{t}^{H} \boldsymbol{\Phi}_{1} \mathbf{G}\mathbf{w}_{i}\right|^2-\left|\tau_K\right|^2\xi_K^2
\end{equation}
\hrulefill
\vspace*{2pt}
\end{figure*}
\vspace{-0.0cm}

Each side of the dual-functional RIS has $M$ passive reconfigurable elements, whose phase shifters can be adjusted by a controller through a dedicated control link. We define the reflection coefficient vector $\bm{\phi}_{1}\triangleq \left[e^{j \theta_{1,1}}, \ldots, e^{j \theta_{1,M}}\right]^T$ and reflection matrix $\bm{\Phi}_{1}\triangleq\operatorname{diag}\left(\bm{\phi}_{1}\right)$, with $\theta_{1,m}$ denoting the phase-shift of the $m$-th element of RIS-1. Similarly, the notations $\bm{\phi}_{2}$ and $\bm{\Phi}_{2}$ are defined in the same way to represent the phase-shift of RIS-2.
We consider a challenging scenario where no direct link exists between the transmitter and the users due to, for example, the blockage of buildings. Since the $k$-th user, $k =1,\cdots, K-1$, is solely served by reflective RIS-1, the received signal at these users can be modeled as
\setcounter{equation}{0}
\begin{equation}
y_{k}= \mathbf{h}_{k}^{H} \bm{\Phi}_{1} \mathbf{G}\sum_{i=1}^{K} \mathbf{w}_{i} s_{i}+n_{k}, ~~~k = 1,\ldots,K-1,\label{eq:y_user1}
\end{equation}
where $s_{i}\in \mathbb{C}, i = 1,\ldots,K$, is the transmitted symbol, $\mathbb{E}\{\left|s_{i}\right|^{2}\}=1$, $\mathbf{w}_{i} \in \mathbb{C}^{N}$ is the beamformer at the BS for the $i$-th user, $\mathbf{G}\in \mathbb{C}^{M\times N}$ and $\mathbf{h}_{k} \in \mathbb{C}^{M}$ denote the channels from BS to RIS-1, from RIS-1 to user-$k$, respectively, and $n_{k} \sim \mathcal{CN}\left(0, \sigma^{2}_k\right)$ denotes the complex additive white Gaussian noise with variance $\sigma^{2}_k$ at user-$k$. We assume that the channel state information (CSI) of all channels involved is perfectly known. The practical CSI acquisition techniques can be found in \cite{swdhz}.
Thus, the signal-to-interference-plus-noise ratio (SINR) of the $k$-th  user can be expressed as
\begin{equation}
\begin{aligned}
\mathrm{SINR}_{k}= \frac{|\mathbf{h}_{k}^{H} \boldsymbol{\Phi}_{1} \mathbf{G}\mathbf{w}_{k}|^2}{\sum_{i \neq k}^{K}\left|\mathbf{h}_{k}^{H} \boldsymbol{\Phi}_{1} \mathbf{G}\mathbf{w}_{i} \right|^{2}+{\sigma_{k}^{2}}},k=1,\ldots,K-1.
\end{aligned}
\end{equation}

Similarly, the received signal at the input of amplifier via the horn antenna-1 can be written as
\begin{equation}
y_\mathrm{r}= \mathbf{g}_\mathrm{t}^{H} \boldsymbol{\Phi}_{1} \mathbf{G}\sum_{i=1}^{K} \mathbf{w}_{i} s_{i}+n_{0}, \label{eq:y_r}
\end{equation}
where $n_{0} \sim \mathcal{CN}\left(0, \sigma_{0}^{2}\right)$ denotes the thermal noise with variance $\sigma_0^{2}$, $\mathbf{g}_\mathrm{t} \in \mathbb{C}^{M}$ denotes the channel from RIS-1 to horn antenna-1. Then, the signal $y_\mathrm{r}$ is amplified and transmitted by RIS-2 towards the $K$-th user, whose received signal can be written as
\begin{equation}
\begin{aligned}
y_{K}=&\sqrt{\beta}\mathbf{h}_{K}^H\boldsymbol{\Phi}_{2} \mathbf{g}_\mathrm{r}\left(\mathbf{g}_\mathrm{t}^{H} \boldsymbol{\Phi}_{1} \mathbf{G}\sum_{i=1}^{K} \mathbf{w}_{i} s_{i}+n_{0}\right)+n_{K}\\
=&\sqrt{\beta}\mathbf{h}_{K}^H\boldsymbol{\Phi}_{2} \mathbf{g}_\mathrm{r} \mathbf{g}_\mathrm{t}^{H} \boldsymbol{\Phi}_{1} \mathbf{G}\sum_{i=1}^{K} \mathbf{w}_{i} s_{i}+\sqrt{\beta}\mathbf{h}_{K}^H\boldsymbol{\Phi}_{2} \mathbf{g}_\mathrm{r}n_{0}+n_{K},\label{eq:y_user2}
\end{aligned}
\end{equation}
where $\mathbf{g}_\mathrm{r} \in \mathbb{C}^{M}$ denotes the channel from horn antenna-2 to RIS-2 and $\beta$ denotes the amplification gain which is assumed to be a constant.
%with the constraint
%\begin{equation}
%\begin{aligned}
%\beta \leq \frac{P_\mathrm{R}}{\left|\mathbf{g}_\mathrm{t}^{H}\boldsymbol{\Phi}_{1}\mathbf{G}\sum_{i = 1}^{K} \mathbf{w}_{i} \right|^{2}+\sigma_{0}^{2}}.
%\end{aligned}
%\end{equation}
Considering that the RISs and the horn antennas are deployed very closely, we use near-field propagation model to describe channels $\mathbf{g}_\mathrm{t}$ and $\mathbf{g}_\mathrm{r}$ \cite{Cheng}, \cite{near-field}, whose details are omitted due to space limitations.
%\cite{Cheng},
Thus, the SINR of the $K$-th user can be expressed as
\begin{equation}
\begin{aligned}
\mathrm{SINR}_{K}=\frac{\beta|\mathbf{h}_{K}^H\boldsymbol{\Phi}_{2} \mathbf{g}_\mathrm{r} \mathbf{g}_\mathrm{t}^{H} \boldsymbol{\Phi}_{1} \mathbf{G}\mathbf{w}_{K}|^2}{\beta\sum_{i \neq K}^{K}\left|\mathbf{h}_{K}^H\boldsymbol{\Phi}_{2} \mathbf{g}_\mathrm{r} \mathbf{g}_\mathrm{t}^{H} \boldsymbol{\Phi}_{1} \mathbf{G}\mathbf{w}_{i}\right|^2+\xi_K^2},
\end{aligned}
\end{equation}
where $\xi_K^2\triangleq \beta\sigma_{0}^{2}|\mathbf{h}_{K}^H\boldsymbol{\Phi}_{2} \mathbf{g}_\mathrm{r}|^2+\sigma_{K}^{2}$ denotes the power of the received noise at the $K$-th user.

This paper aims at maximizing the sum-rate of the MU-MISO downlink system by jointly designing the BS beamforming $\mathbf{w}_k, \forall k$, and RIS reflection matrices $\boldsymbol{\Phi}_{1}, \boldsymbol{\Phi}_{2}$ of the dual-functional RIS, subject to the following constraints: i) Transmit power at the BS is less than or equal to power budget $P_\mathrm{T}$, i.e., $\sum_{k=1}^{K}\left\|\mathbf{w}_{k}\right\|^2 \leq P_\mathrm{T}$; ii) Constant magnitude of reflection coefficients, i.e. $|\bm{\phi}_{1}(m)|=1, |\bm{\phi}_{2}(m)|=1, \forall m$. Therefore, the corresponding optimization problem can be formulated as
\vspace{-0.0cm}
%\begin{small}
\begin{subequations}\label{eq:problem1}
\begin{align}
\mathrm{(P1)} \max_{\boldsymbol{\Phi}_{1},\boldsymbol{\Phi}_{2},\mathbf{w}_{k}} & \sum_{k=1}^{K} \log _{2}\left(1+\operatorname{SINR}_{k}\right) \\
\mathrm{s.t.} ~~~
%\beta \leq \frac{P_\mathrm{R}}{\left|\mathbf{g}_\mathrm{t}^{H}\boldsymbol{\Phi}_{1}\mathbf{G}\sum_{i = 1}^{K} \mathbf{w}_{i} \right|^{2}+\sigma_{0}^{2}}, \label{eq:problem11}\\
& \sum_{k=1}^{K}\left\|\mathbf{w}_{k}\right\|^2 \leq P_\mathrm{T}, \label{eq:problem12}\\
%&\bm{\Phi}_{1}= \mathrm{diag}\left(\bm{\phi}_{1}\right), \bm{\Phi}_{2}=\mathrm{diag}\left(\bm{\phi}_{2}\right), \label{eq:problem13}\\
& |\bm{\phi}_{1}(m)|=1, ~|\bm{\phi}_{2}(m)|=1,~\forall m. \label{eq:problem15}
\end{align}
\end{subequations}
%\end{small}
%where $P_\mathrm{T}$ is the total transmit power budget at the BS.
Obviously, the optimization problem \eqref{eq:problem1} is non-convex and difficult to solve due to the unit-modulus constraints and the coupling between variables. To tackle the difficulties in \eqref{eq:problem1}, in the following we propose an iterative BS beamforming and RIS reflection design algorithm based on fractional programming (FP) theory.

\vspace{-0.0 cm}

\section{Joint Beamforming and RIS Design}
%\vspace{-0.1 cm}
In this section, we first equivalently transform the original problem into a more tractable form based on the theory of FP. With the transformed objective function, BS beamforming and RIS reflection matrices are designed iteratively.

%\vspace{-0.1 cm}
\vspace{-0.0cm}
\subsection{FP-Based Transformation of Objective Function}
%\vspace{-0.1cm}

The objective in \eqref{eq:problem1} is a typical function of multiple fractional parameters, which is usually difficult to solve. Motivated by \cite{Shen1}, \cite{Shen2} which use FP to solve this multiple-ratio problem, we attempt to equivalently transform the original problem into a more tractable form by extracting the ratio terms SINR$_k$, $k = 1,\ldots,K,$ from the logarithmic function. Based on Lagrangian dual transform \cite{Shen1}, the optimization problem \eqref{eq:problem1} is equivalent to
\begin{equation}\label{eq:problem21}
\begin{aligned}
\mathrm{(P2)} \max_{\boldsymbol{\Phi}_{1},\boldsymbol{\Phi}_{2},\mathbf{w}_{k}, \bm{\gamma}} & f_\mathrm{R}(\boldsymbol{\Phi}_{1},\boldsymbol{\Phi}_{2},\mathbf{w}_{k},\bm{\gamma}) \\
\mathrm{s.t.} ~~~~~&\eqref{eq:problem12},\eqref{eq:problem15},
%\beta \leq \frac{P_\mathrm{R}}{\left|\mathbf{g}_\mathrm{t}^{H}\boldsymbol{\Phi}_{1}\mathbf{G}\sum_{i = 1}^{K} \mathbf{w}_{i} \right|^{2}+\sigma_{0}^{2}}, \label{eq:problem11}\\
%& \sum_{k=1}^{K}\left\|\mathbf{w}_{k}\right\|^2 \leq P_\mathrm{T}, \label{eq:problem22}\\
%&\bm{\Phi}_{1}= \mathrm{diag}\left(\bm{\phi}_{1}\right), \bm{\Phi}_{2}=\mathrm{diag}\left(\bm{\phi}_{2}\right), \label{eq:problem13}\\
%& |\bm{\phi}_{1}(i)|=1, ~|\bm{\phi}_{2}(i)|=1,~\forall i, \label{eq:problem25}
\end{aligned}
\end{equation}
where the objective function in \eqref{eq:problem21} is
\begin{equation} \label{eq:f1}
\begin{aligned}
f_\mathrm{R}(\boldsymbol{\Phi}_{1},\boldsymbol{\Phi}_{2},\mathbf{w}_{k},\bm{\gamma}&)
=\sum_{k=1}^{K} \log _{2}\left(1+\gamma_{k}\right)-\sum_{k=1}^{K}\gamma_{k}\\
+&\sum_{k=1}^{K-1}\frac{\left(1+\gamma_{k}\right)|\mathbf{h}_{k}^{H} \boldsymbol{\Phi}_{1} \mathbf{G}\mathbf{w}_{k}|^2}{\sum_{i=1}^{K}\left|\mathbf{h}_{k}^{H} \boldsymbol{\Phi}_{1} \mathbf{G}\mathbf{w}_{i} \right|^{2}+{\sigma_{k}^{2}}}\\
+&\frac{\beta\left(1+\gamma_{K}\right)|\mathbf{h}_{K}^H\boldsymbol{\Phi}_{2} \mathbf{g}_\mathrm{r} \mathbf{g}_\mathrm{t}^{H} \boldsymbol{\Phi}_{1} \mathbf{G}\mathbf{w}_{K}|^2}{\beta\sum_{i=1}^{K}\left|\mathbf{h}_{K}^H\boldsymbol{\Phi}_{2} \mathbf{g}_\mathrm{r} \mathbf{g}_\mathrm{t}^{H} \boldsymbol{\Phi}_{1} \mathbf{G}\mathbf{w}_{i}\right|^2+\xi_K^2},
\end{aligned}
\end{equation}
%when each element of the auxiliary variable vector $\bm{\gamma}\triangleq[\gamma_1,\ldots,\gamma_K]^T$ has the following optimal value
%\begin{equation} \label{eq:rrr}
%\gamma_{k}^{\star}=\left\{
%\begin{aligned}
%&\frac{|\mathbf{h}_{k}^{H} \boldsymbol{\Phi}_{1} \mathbf{G}\mathbf{w}_{k}|^2}{\sum_{i \neq k}^{K}\left|\mathbf{h}_{k}^{H} \boldsymbol{\Phi}_{1} \mathbf{G}\mathbf{w}_{i} \right|^{2}+{\sigma_{k}^{2}}},k=1,\ldots,K-1,\\
%&\frac{\beta|\mathbf{h}_{k}^H\boldsymbol{\Phi}_{2} \mathbf{g}_\mathrm{r} \mathbf{g}_\mathrm{t}^{H} \boldsymbol{\Phi}_{1} \mathbf{G}\mathbf{w}_{k}|^2}{\beta\sum_{i \neq k}^{K}\left|\mathbf{h}_{k}^H\boldsymbol{\Phi}_{2} \mathbf{g}_\mathrm{r} \mathbf{g}_\mathrm{t}^{H} \boldsymbol{\Phi}_{1} \mathbf{G}\mathbf{w}_{i}\right|^2+\xi},k=K.
%\end{aligned}
%\right.
%\end{equation}
%
%
%Therefore, the original objective in \eqref{eq:problem1} can be firstly transformed into an equivalent
%function as shown in \eqref{eq:reproblem}.
and $\bm{\gamma} \triangleq [\gamma_1,\ldots,\gamma_K]^T$ is an auxiliary variable vector. Unfortunately, optimization problem \eqref{eq:problem21} is still intractable due to the complicated form of the sum of $K$ fractional terms. Next, we apply quadratic transform \cite{Shen2} on the fractional term to transform it into a solvable formula by introducing another auxiliary variable vector $\bm{\tau}\triangleq[\tau_1,\ldots,\tau_K]^T$. Then the optimization problem \eqref{eq:problem21} can be transformed into
\setcounter{equation}{8}
\begin{equation}
\begin{aligned}
\mathrm{(P3)} \max_{\boldsymbol{\Phi}_{1},\boldsymbol{\Phi}_{2},\mathbf{w}_{k}, \bm{\gamma}, \bm{\tau}} & g_\mathrm{R}(\boldsymbol{\Phi}_{1},\boldsymbol{\Phi}_{2},\mathbf{w}_{k},\bm{\gamma},\bm{\tau}) \label{eq:problem31}\\
\mathrm{s.t.} ~~~~~\eqref{eq:problem12},\eqref{eq:problem15},
%\beta \leq \frac{P_\mathrm{R}}{\left|\mathbf{g}_\mathrm{t}^{H}\boldsymbol{\Phi}_{1}\mathbf{G}\sum_{i = 1}^{K} \mathbf{w}_{i} \right|^{2}+\sigma_{0}^{2}}, \label{eq:problem11}\\
%& \sum_{k=1}^{K}\left\|\mathbf{w}_{k}\right\|^2 \leq P_\mathrm{T}, \label{eq:problem22}\\
%&\bm{\Phi}_{1}= \mathrm{diag}\left(\bm{\phi}_{1}\right), \bm{\Phi}_{2}=\mathrm{diag}\left(\bm{\phi}_{2}\right), \label{eq:problem13}\\
%& |\bm{\phi}_{1}(i)|=1, ~|\bm{\phi}_{2}(i)|=1,~\forall i, \label{eq:problem25}\label{eq:problem3}
\end{aligned}
\end{equation}
where the objective function $g_\mathrm{R}(\boldsymbol{\Phi}_{1},\boldsymbol{\Phi}_{2},\mathbf{w}_{k},\bm{\gamma},\bm{\tau})$ is formulated as \eqref{eq:reR2} presented at the top of this page.

To efficiently solve the above problem, we adopt the block coordinate ascent (BCA) methodology to alternatively update each block of variables while keep others being fixed. The sub-problems of updating each block will be specified in details in the following.
%To efficiently solve this problem, we propose to iteratively update the variables $\boldsymbol{\Phi}_{1}, \boldsymbol{\Phi}_{2}, \mathbf{w}_{k}, \bm{\gamma}$ and $\bm{\tau}$ to find the conditionally optimal solution of one variable matrix/vector given others at each iteration.
%%While the conditionally optimal
%%$\bm{\gamma}$ and $\bm{\tau}$ are already shown in \eqref{eq:rrr}, \eqref{eq:yyy1} and \eqref{eq:yyy2}, respectively,
%In the following subsection, we will discuss the solution for each variable vector/matrix in details.
%\vspace{-0.2cm}
\subsection{Update Auxiliary Variable Vectors}
1) Update $\bm{\gamma}$

When $\boldsymbol{\Phi}_{1}, \boldsymbol{\Phi}_{2}, \mathbf{w}_{k}$ and $\bm{\tau}$ are held fixed, the objective function $g_\mathrm{R}$ in \eqref{eq:problem31} is a concave differentiable
function with respect to the variables $\gamma_{k}$, $\forall k$. By solving $\frac{\partial g_\mathrm{R}}{\partial \gamma_{k}}=0$, the optimal ${\gamma}_k^{\star}$ can be found in a closed-form as
\setcounter{equation}{10}
\begin{equation} \label{eq:rrr}
\gamma_{k}^{\star}=\left\{
\begin{aligned}
&\frac{|\mathbf{h}_{k}^{H} \boldsymbol{\Phi}_{1} \mathbf{G}\mathbf{w}_{k}|^2}{\sum_{i \neq k}^{K}\left|\mathbf{h}_{k}^{H} \boldsymbol{\Phi}_{1} \mathbf{G}\mathbf{w}_{i} \right|^{2}+{\sigma_{k}^{2}}},k=1,\ldots,K-1,\\
&\frac{\beta|\mathbf{h}_{k}^H\boldsymbol{\Phi}_{2} \mathbf{g}_\mathrm{r} \mathbf{g}_\mathrm{t}^{H} \boldsymbol{\Phi}_{1} \mathbf{G}\mathbf{w}_{k}|^2}{\beta\sum_{i \neq k}^{K}\left|\mathbf{h}_{k}^H\boldsymbol{\Phi}_{2} \mathbf{g}_\mathrm{r} \mathbf{g}_\mathrm{t}^{H} \boldsymbol{\Phi}_{1} \mathbf{G}\mathbf{w}_{i}\right|^2+\xi_K^2},k=K.
\end{aligned}
\right.
\end{equation}
%whose proof is omitted due to space limitations.

\vspace{0.1cm}
2) Update $\bm{\tau}$

Similarly, given $\boldsymbol{\Phi}_{1}, \boldsymbol{\Phi}_{2}, \mathbf{w}_{k}$ and $\bm{\gamma}$, $g_\mathrm{R}$ in \eqref{eq:problem31} is a concave differentiable function with respect to
$\tau_{k}$, $\forall k$. The optimal variable $\tau_{k}^{\star}$ can be obtained by setting $\frac{\partial g_\mathrm{R}}{\partial \tau_{k}}=0$ and has the following optimal value
\begin{equation} \label{eq:ttt}
\tau_{k}^{\star}=\left\{
\begin{aligned}
&\frac{\sqrt{1+\gamma_{k}}\mathbf{h}_{k}^{H} \boldsymbol{\Phi}_{1} \mathbf{G}\mathbf{w}_{k}}{\sum_{i=1}^{K}\left|\mathbf{h}_{k}^{H} \boldsymbol{\Phi}_{1} \mathbf{G}\mathbf{w}_{i} \right|^{2}+{\sigma_{k}^{2}}},k=1,\ldots,K-1,\\
&\frac{\sqrt{\beta\left(1+\gamma_{k}\right)}\mathbf{h}_{k}^H\boldsymbol{\Phi}_{2} \mathbf{g}_\mathrm{r} \mathbf{g}_\mathrm{t}^{H} \boldsymbol{\Phi}_{1} \mathbf{G}\mathbf{w}_{k}}{\beta\sum_{i=1 }^{K}\left|\mathbf{h}_{k}^H\boldsymbol{\Phi}_{2} \mathbf{g}_\mathrm{r} \mathbf{g}_\mathrm{t}^{H} \boldsymbol{\Phi}_{1} \mathbf{G}\mathbf{w}_{i}\right|^2+\xi_K^2},k=K.
\end{aligned}
\right.
\end{equation}

%whose proof is also omitted due to space limitations.
%\vspace{-0.3cm}
\subsection{Update BS Beamforming}

With other variables being fixed, the update of beamforming $\mathbf{w}_{k}$ is transformed into solving the following problem
%When $\bm{\gamma}, \bm{\tau}$ and the RIS reflections $\boldsymbol{\Phi}_{1},\boldsymbol{\Phi}_{2}$ are fixed, the problem of conditionally optimal BS beamforming design can be presented as follows:
%\setcounter{equation}{11}
\begin{subequations}\label{eq:BS}
\begin{align}
\mathrm{(P4)}~\max_{\mathbf{w}_{k}}~& g_\mathrm{R}(\boldsymbol{\Phi}_{1},\boldsymbol{\Phi}_{2},\mathbf{w}_{k},\bm{\gamma},\bm{\tau}) \\
\mathrm{s.t.}~~ & \sum_{k = 1}^{K}\left\|\mathbf{w}_{k}\right\|^2 \leq P_\mathrm{T},\label{eq:BSp}
\end{align}
\end{subequations}
which is a convex problem. Fortunately, the optimal solution of problem \eqref{eq:BS} can be obtained in an analytic form via checking its optimality conditions. By introducing a multiplier $\mu$ for the power constraint in \eqref{eq:BSp}, we can form a Lagrangian
function as
%\vspace{-0.4cm}
\begin{equation}
\begin{aligned}
\mathcal{L}= g_\mathrm{R}(\boldsymbol{\Phi}_{1},\boldsymbol{\Phi}_{2},\mathbf{w}_{k},\bm{\gamma},\bm{\tau})+\mu\left(P_\mathrm{T}-\sum_{k = 1}^{K}\left\|\mathbf{w}_{k}\right\|^2\right).
\end{aligned}
\end{equation}
%Therefore, the problem \eqref{eq:BS} can be reformulated as
%\begin{equation}\label{eq:BS2}
%\begin{aligned}
%\{\mathbf{w}_{k}^{\star},\mu^{\star}\} =\arg\max L_{\delta}.
%\end{aligned}
%\end{equation}
Then, the optimal solution of BS beamforming $\mathbf{w}_k$ can be determined by setting the partial derivative of $\mathcal{L}$ with respect to $\mathbf{w}_k$ and $\mu$ to zero, i.e.
\begin{equation}
%\left\{\begin{array}
%\frac{\partial L_{\delta}}{\partial \mathbf{f}_{\mathrm{BB}, k}}=\mathbf{0}, \quad \forall k \\
\left\{\begin{aligned}
&\frac{\partial \mathcal{L}}{\partial \mathbf{w}_k}= 0, ~~ \forall k, \\
&\frac{\partial \mathcal{L}}{\partial \mu}=0,\\
\end{aligned}\right.
%\frac{\partial L_{\delta}}{\partial \mathbf{f}_{\mathrm{BB}, k}}=\mathbf{0}, \quad \forall k \\
%\frac{\partial L_{\delta}}{\partial \mu}=0
%\end{array}\right.
\end{equation}
which yields the optimal BS beamforming as
\setcounter{equation}{15}
 \begin{equation} \label{eq:w}
\mathbf{w}_k^{\star} =
%\left\{\begin{aligned}
\left(\mathbf{A}+\mu^{\star}\mathbf{I}\right)^{-1}\sqrt{1+\gamma_{k}}\widetilde{\mathbf{h}}_k,~~\forall k,
%&\left(\mathbf{A}+\mu^{\star}\mathbf{I}\right)^{-1}\sqrt{\left(1+\gamma_{k}\right)}\widetilde{\mathbf{h}}_2,k=K,\\
%\end{aligned}\right.
\end{equation}
with
\begin{equation}
\widetilde{\mathbf{h}}_k^H \triangleq
\left\{\begin{aligned}
&\tau_k^\ast\mathbf{h}_{k}^H\boldsymbol{\Phi}_{1}\mathbf{G}, ~~k = 1,\ldots,K-1,\\
&\sqrt{\beta}\tau_k^\ast\mathbf{h}_{k}^H\boldsymbol{\Phi}_{2}\mathbf{g}_\mathrm{r}\mathbf{g}_\mathrm{t}^H\boldsymbol{\Phi}_{1}\mathbf{G},k = K,
\end{aligned}\right.
\end{equation}
\begin{equation}
\begin{aligned}
\mathbf{A} \triangleq  \sum_{k=1}^{K}\widetilde{\mathbf{h}}_k\widetilde{\mathbf{h}}_k^H.
%%\left|\tau_j\right|^2\mathbf{G}^H\boldsymbol{\Phi}_{1}^H\mathbf{h}_{j}\mathbf{h}_{j}^{H} \boldsymbol{\Phi}_{1} \mathbf{G}\\
%%+&\beta\left|\tau_K\right|^2\mathbf{G}^H\boldsymbol{\Phi}_{1} ^H\mathbf{g}_\mathrm{t}\mathbf{g}_\mathrm{r}^H\boldsymbol{\Phi}_{2}^H\mathbf{h}_{K}\mathbf{h}_{K}^H\boldsymbol{\Phi}_{2} \mathbf{g}_\mathrm{r} \mathbf{g}_\mathrm{t}^{H} \boldsymbol{\Phi}_{1} \mathbf{G},
\end{aligned}
\end{equation}
The optimal BS beamforming $\mathbf{w}_{k}$ can be easily obtained after having the optimal multiplier $\mu^{\star}$ by a bisection search.

\subsection{Update RIS Reflection}

Given $\bm{\gamma}$, $\bm{\tau}$ and BS beamforming $\mathbf{w}_k, \forall k$, the sub-problem with respect to the RIS reflection matrices $\bm{\Phi}_{1}$ and $\bm{\Phi}_{2}$ can be presented as:
\begin{subequations}  \label{p:Psi}
\begin{align}
\mathrm{(P5)}~~ \max_{\bm{\Phi}_{1},\bm{\Phi}_{2}}  ~~& g_\mathrm{R}(\boldsymbol{\Phi}_{1},\boldsymbol{\Phi}_{2},\mathbf{w}_{k},\bm{\gamma},\bm{\tau}) \\
\mathrm{s.t.} ~~~~
%&\eqref{eq:problem13}-\eqref{eq:problem15}
%&\bm{\Phi}_{1}= \mathrm{diag}\left(\bm{\phi}_{1}\right), \bm{\Phi}_{2}=\mathrm{diag}\left(\bm{\phi}_{2}\right),\\
& |\bm{\phi}_{1}(m)|=1,~~ |\bm{\phi}_{2}(m)|=1,~~\forall m. \label{eq:nuit2}
\end{align}
\end{subequations}
%We define $\bm{\phi}\triangleq [\bm{\phi}_{1};\bm{\phi}_{2}]$, $\mathbf{H}_1\triangleq \left[\mathrm{diag}\left(\mathbf{h}_{k}^{H}\right); \mathbf{0}\right]$, $\mathbf{H}_2\triangleq \left[ \mathbf{0}; \mathrm{diag}\left(\mathbf{h}_{k}^{H}\right)\right]$, and $\mathbf{G}_{\mathrm{r},1}\triangleq \left[ \mathrm{diag}\left(\mathbf{g}_\mathrm{t}^{H}; \mathbf{0}\right)\right]$ and the objective function in \eqref{p:Psi} can be rearranged as
%\begin{equation}
%\begin{aligned}
%\delta = \sum_{k=1}^{K-1}\left(2\sqrt{1+r_{k}}\Re\left\{y_k^{\star}\bm{\phi}^{H}\mathbf{H}_1 \mathbf{G}\mathbf{w}_{k}\right\}-\left|y_k\right|^2\sum_{i=1}^{K}\left|\bm{\phi}^{H} \mathbf{H}_1 \mathbf{G}\mathbf{w}_{i} \right|^{2}\right)\\
%+
%2\sqrt{\beta\left(1+r_{K}\right)}\Re\left\{y_K^{\star}\bm{\phi}^{H}\mathbf{H}_2 \mathbf{g}_\mathrm{r} \bm{\phi}^{H}\mathbf{G}_{\mathrm{t},1} \mathbf{G}\mathbf{w}_{K}\right\}\\
%-\left|y_K\right|^2\beta\sum_{i=1}^{K}\left|\bm{\phi}^{H}\mathbf{H}_2 \mathbf{g}_\mathrm{r}\bm{\phi}^{H}\mathbf{G}_{\mathrm{r},1}\mathbf{G}\mathbf{w}_{i}\right|^2
%\end{aligned}
%\end{equation}
Considering that $\boldsymbol{\Phi}_{1}$ and $\boldsymbol{\Phi}_{2}$ are coupled and jointly yield a nonconvex objective, we turn to separately optimize either of them with the other being fixed.
%It can be seen that $\boldsymbol{\Phi}_{1}$ and $\boldsymbol{\Phi}_{2}$ are coupled, which means it is difficult to obtain $\boldsymbol{\Phi}_{1}$ and $\boldsymbol{\Phi}_{2}$ together. Hence, we attempt to solve $\boldsymbol{\Phi}_{1}$ and $\boldsymbol{\Phi}_{2}$ iteratively to reduce the complexity and solve $\boldsymbol{\Phi}_{1}$ firstly with fixed $\boldsymbol{\Phi}_{2}$.
When $\boldsymbol{\Phi}_{2}$ is given, by defining
%\begin{small}
\begin{equation}
\mathbf{r}_{k,i}^H\triangleq
\left\{\begin{aligned}
&\tau_k^{\ast}\mathbf{h}_{k}^H\mathrm{diag}\left(\mathbf{G}\mathbf{w}_{i}\right),~k=1,\ldots,K-1,~\forall i,\\
&\sqrt{\beta}\tau_k^{\ast}\mathbf{h}_{k}^H\boldsymbol{\Phi}_{2}\mathbf{g}_\mathrm{r} \mathbf{g}_\mathrm{t}^{H} \mathrm{diag}\left( \mathbf{G}\mathbf{w}_{i}\right),k=K,\forall i,
\end{aligned}
\right.
\end{equation}
%\end{small}
\begin{align}
\widetilde{\mathbf{R}}&\triangleq \sum_{k=1}^{K}\sum_{i=1}^{K}\mathbf{r}_{k,i}\mathbf{r}_{k,i}^H,\\
\mathbf{d}&\triangleq
\sum_{k=1}^{K}\sqrt{\left(1+\gamma_{k}\right)}\mathbf{r}_{k,k},
\end{align}
and removing constants, the optimization of $\boldsymbol{\phi}_{1}$ is equivalent to solving the following problem
\begin{subequations}  \label{p:Psi1}
\begin{align}
\mathrm{(P6)}\min_{\boldsymbol{\phi}_{1}} ~
&-\sum_{k=1}^{K}\left(2\sqrt{1+\gamma_{k}}\Re\left\{\mathbf{r}_{k,k}^H\bm{\phi}_{1} \right\}-\sum_{i=1}^{K}\left|\mathbf{r}_{k,i}^H\bm{\phi}_{1} \right|^{2}\right)\nonumber \\
&=\bm{\phi}_{1}^H\widetilde{\mathbf{R}}\bm{\phi}_{1}-2\Re\left\{\bm{\phi}_{1}^H\mathbf{d}\right\}\\
%&=\boldsymbol{\phi}_{2}^H\widetilde{\mathbf{V}}\boldsymbol{\phi}_{2}-2\Re\left\{\boldsymbol{\phi}_{2}^H\mathbf{b}\right\}-{a_1}\\
 %\stackrel{(a)}{=}\boldsymbol{\phi}_{2}^H\widetilde{\mathbf{V}}\boldsymbol{\phi}_{2}-2\Re\left\{\boldsymbol{\phi}_{2}^H\mathbf{b}\right\}\\
\mathrm{s.t.}~ %~&\bm{\Phi}_{2}=\mathrm{diag}\left(\bm{\phi}_{2}\right),\\
&|\bm{\phi}_{1}(m)|=1,~m=1,\cdots,M.
\end{align}
\end{subequations}
Problem \eqref{p:Psi1} is still difficult to solve due to the constant modulus constraint. To effectively solve this problem, we adopt the majorization-minimization (MM) algorithm to perform the updating. The essential idea of MM framework is to sequentially surrogate the original objective function with a tightly upper-bounded approximation, which is designed to render the transformed problem more tractable \cite{Yang}. Given the update value $\boldsymbol{\phi}_{1}^t$ in the previous iteration, we
propose to majorize the objective function of problem \eqref{p:Psi1} via executing its second-order Taylor expansion at the point of $\boldsymbol{\phi}_{1}^t$ followed by replacing the associated quadratic term with an upper bound. Specifically, the quadratic term in \eqref{p:Psi1} is majorized by
\vspace{0.1cm}
\begin{equation}
\begin{aligned}
\bm{\phi}_1^{H} \widetilde{\mathbf{R}}\bm{\phi}_1 \leq &\bm{\phi}_1^{H} \bm{\Lambda} \bm{\phi}_1 -2\Re\left\{\bm{\phi}_1^{H}(\bm{\Lambda}-\widetilde{\mathbf{R}}) \bm{\phi}_1^{t}\right\} \\
&+\left(\bm{\phi}_1^{t}\right)^{\mathrm{H}}(\bm{\Lambda}-\widetilde{\mathbf{R}})\bm{\phi}_1^{t},
\end{aligned}
\end{equation}
where $\bm{\Lambda} \triangleq \lambda_{\max}\mathbf{I}_{M}$ and $\lambda_{\max}$ is the maximum eigenvalue of $\widetilde{\mathbf{R}}$. The subproblem to be solved at the $(t+1)$-th iteration is given by
\vspace{0.1cm}
\begin{subequations}  \label{p:rePsi1}
\begin{align}
\min_{\bm{\phi}_{1}}~~&\bm{\phi}_1^{H} \bm{\Lambda} \bm{\phi}_1 -2\Re\left\{\bm{\phi}_1^{H}(\bm{\Lambda}-\widetilde{\mathbf{R}}) \bm{\phi}_1^{t}\right\}\nonumber \\
&+\left(\bm{\phi}_1^{t}\right)^{\mathrm{H}}(\bm{\Lambda}-\widetilde{\mathbf{R}})\bm{\phi}_1^{t}-2\Re\left\{\boldsymbol{\phi}_{1}^H\mathbf{d}\right\}\\
\mathrm{s.t.} ~~%&\bm{\Phi}_{2}=\mathrm{diag}\left(\bm{\phi}_{2}\right),\\
&|\bm{\phi}_{1}(m)|=1,~m=1,\cdots,M.
\end{align}
\end{subequations}
Since $\bm{\phi}_{1}^H\bm{\phi}_{1} = M$, we have $\bm{\phi}_{1}^H\bm{\Lambda}\bm{\phi}_{1} = M\lambda_{\max}$, which is a constant. By removing the constants, problem \eqref{p:rePsi1} can be rewritten as follows
\vspace{0.1cm}
\begin{subequations}  \label{p:lastPsi1}
\begin{align}
\max_{\bm{\phi}_{1}}~~&2\Re\left\{\bm{\phi}_1^H\mathbf{p}^t\right\}\\
\mathrm{s.t.} ~~%&\bm{\Phi}_{2}=\mathrm{diag}\left(\bm{\phi}_{2}\right),\\
&|\bm{\phi}_{1}(m)|=1,~m=1,\cdots,M,
\end{align}
\end{subequations}
where $\mathbf{p}^t \triangleq (\lambda_{\max}\mathbf{I}_{M}-\widetilde{\mathbf{R}})\bm{\phi}_1^t+\mathbf{d}$. Thus, the optimal solution of problem \eqref{p:lastPsi1} is given by
\vspace{0.1cm}
\begin{equation}
\bm{\phi}_1^{t+1}=e^{j \angle\mathbf{p}^{t}}. \label{eq:psi1}
\end{equation}
%where $\arg ()$ is an operation to obtain the angle of a vector.

After obtaining $\boldsymbol{\phi}_{1}$, we can solve $\boldsymbol{\phi}_{2}$ by repeating the same procedure. By defining
\begin{align}
\mathbf{v}_{K,k}^H &\triangleq \sqrt{\beta}\tau_K^{\ast}\mathbf{h}_{K}^H\mathrm{diag}\left(\mathbf{g}_\mathrm{r} \mathbf{g}_\mathrm{t}^{H} \boldsymbol{\Phi}_{1} \mathbf{G}\mathbf{w}_{k}\right),\forall k,\\
\widetilde{\mathbf{V}}&\triangleq \sum_{k=1}^{K}\mathbf{v}_{K,k}\mathbf{v}_{K,k}^H,\\
\mathbf{b}&\triangleq
\sqrt{\left(1+\gamma_{K}\right)}\mathbf{v}_{K,K},
\end{align}
%\begin{small}
%\begin{equation}
%\begin{aligned}
%{a_1} &\triangleq \sum_{k=1}^{K-1}\left(2\sqrt{1+\gamma_{k}}\Re\left\{\widetilde{\mathbf{h}}_{k}^{H} \mathbf{w}_{k}\right\}-\sum_{i=1}^{K}\left|\widetilde{\mathbf{h}}_{k}^{H} \mathbf{w}_{i} \right|^{2}\right),
%\end{aligned}
%\end{equation}
%\end{small}
%\begin{equation}
%\begin{aligned}
%\widetilde{\mathbf{h}}_{k}^H&\triangleq \tau_k^{\ast}\mathbf{h}_{k}^{H} \boldsymbol{\Phi}_{1} \mathbf{G},
%\end{aligned}
%\end{equation}
%By defining $\mathbf{v}_{r,K,k} \triangleq \left[\mathbf{h}_{K}^H\mathrm{diag}\left(\mathbf{g}_\mathrm{r} \mathbf{g}_\mathrm{t}^{H} \boldsymbol{\Phi}_{1} \mathbf{G}\mathbf{w}_{k}\right)\right]^H,\forall k$ and $\overline{a_1} \triangleq \sum_{k=1}^{K-1}\left(2\sqrt{1+\gamma_{k}}\Re\left\{\tau_k^{\ast}\mathbf{h}_{k}^{H} \boldsymbol{\Phi}_{1} \mathbf{G}\mathbf{w}_{k}\right\}-\left|\tau_k\right|^2\sum_{i=1}^{K}\left|\mathbf{h}_{k}^{H} \boldsymbol{\Phi}_{1} \mathbf{G}\mathbf{w}_{i} \right|^{2}\right)$,
% %as \eqref{eq:v1}-\eqref{eq:a1},
and removing the other constant terms unrelated to $\boldsymbol{\phi}_{2}$, problem \eqref{p:Psi} can be rearranged as
\begin{subequations}  \label{p:Psi2}
\begin{align}
\mathrm{(P7)}\min_{\boldsymbol{\phi}_{2}} &-2\sqrt{\left(1+\gamma_{K}\right)}\Re\left\{\mathbf{v}_{K,K}^H\boldsymbol{\phi}_{2}  \nonumber \right\}+\sum_{i=1}^{K}\left|\mathbf{v}_{K,i}^H\boldsymbol{\phi}_{2}\right|^2\\
&=\boldsymbol{\phi}_{2}^H\widetilde{\mathbf{V}}\boldsymbol{\phi}_{2}-2\Re\left\{\boldsymbol{\phi}_{2}^H\mathbf{b}\right\}\\
 %\stackrel{(a)}{=}\boldsymbol{\phi}_{2}^H\widetilde{\mathbf{V}}\boldsymbol{\phi}_{2}-2\Re\left\{\boldsymbol{\phi}_{2}^H\mathbf{b}\right\}\\
\mathrm{s.t.}~ %~&\bm{\Phi}_{2}=\mathrm{diag}\left(\bm{\phi}_{2}\right),\\
&|\bm{\phi}_{2}(m)|=1,~m=1,\cdots,M.
\end{align}
\end{subequations}
Following the identical argument as before, the optimal solution is given by
\begin{equation}
\bm{\phi}_2^{t+1}=e^{j \angle\mathbf{q}^{t}}, \label{eq:psi2}
\end{equation}
where $\mathbf{q}^t \triangleq (\zeta_{\max}\mathbf{I}_{M}-\widetilde{\mathbf{V}})\bm{\phi}_2^t+\mathbf{b}$ and $\zeta_{\max}$ is the maximum eigenvalue of $\widetilde{\mathbf{V}}$.
%
%By defining $\mathbf{v}_{r,K,k} \triangleq \left[\mathbf{h}_{K}^H\mathrm{diag}\left(\mathbf{g}_\mathrm{r} \mathbf{g}_\mathrm{t}^{H} \boldsymbol{\Phi}_{1} \mathbf{G}\mathbf{w}_{k}\right)\right]^H,\forall k$ and $\overline{a_1} \triangleq \sum_{k=1}^{K-1}\left(2\sqrt{1+\gamma_{k}}\Re\left\{\tau_k^{\ast}\mathbf{h}_{k}^{H} \boldsymbol{\Phi}_{1} \mathbf{G}\mathbf{w}_{k}\right\}-\left|\tau_k\right|^2\sum_{i=1}^{K}\left|\mathbf{h}_{k}^{H} \boldsymbol{\Phi}_{1} \mathbf{G}\mathbf{w}_{i} \right|^{2}\right)$,
% %as \eqref{eq:v1}-\eqref{eq:a1},
%and removing the other constant terms unrelated to $\boldsymbol{\phi}_{2}$, problem \eqref{p:Psi} can be rearranged as
%\begin{subequations}  \label{p:Psi2}
%\begin{align}
%\mathrm{(P7)}\min_{\boldsymbol{\phi}_{2}} &-2\sqrt{\left(1+\gamma_{K}\right)}\Re\left\{\mathbf{v}_{K,K}^H\boldsymbol{\phi}_{2}  \nonumber \right\}+\sum_{i=1}^{K}\left|\mathbf{v}_{K,i}^H\boldsymbol{\phi}_{2}\right|^2\\
%&=\boldsymbol{\phi}_{2}^H\widetilde{\mathbf{V}}\boldsymbol{\phi}_{2}-2\Re\left\{\boldsymbol{\phi}_{2}^H\mathbf{b}\right\}\\
% %\stackrel{(a)}{=}\boldsymbol{\phi}_{2}^H\widetilde{\mathbf{V}}\boldsymbol{\phi}_{2}-2\Re\left\{\boldsymbol{\phi}_{2}^H\mathbf{b}\right\}\\
%\mathrm{s.t.}~ %~&\bm{\Phi}_{2}=\mathrm{diag}\left(\bm{\phi}_{2}\right),\\
%&|\bm{\phi}_{2}(m)|=1,~m=1,\cdots,M.
%\end{align}
%\end{subequations}

If low-resolution phase shifters are employed to realize the RIS, the corresponding discrete reflection coefficient can be obtained using a simple quantization operation
\begin{equation} \label{eq:qpsi2}
\begin{aligned}
\widetilde{\bm{\phi}}_{1(2)}^{t+1}=\exp \left\{j\left\langle\frac{\arg\left(\mathbf{p}^{t}(\mathbf{q}^{t})\right)}{\Delta}\right\rangle \times \Delta\right\},
\end{aligned}
\end{equation}
%\begin{equation} \label{eq:qpsi1}
%\begin{aligned}
%\widetilde{\bm{\phi}}_{1}^{t+1}=\exp \left\{j\left\langle\frac{\arg\left(\mathbf{p}^{t}\right)}{\Delta}\right\rangle \times \Delta\right\},
%\end{aligned}
%\end{equation}
where $\langle \cdot \rangle$ denotes the rounding operation, and $\Delta \triangleq 2 \pi / 2^{b}$ is the angle resolution controlled by $b$ bits.

%If low-resolution phase shifters are employed,
%The quantization operation is also applied, i.e.
%\begin{equation} \label{eq:qpsi2}
%\begin{aligned}
%\widetilde{\bm{\phi}}_{2}^{t+1}=\exp \left\{j\left\langle\frac{\arg\left(\mathbf{q}^{t}\right)}{\Delta}\right\rangle \times \Delta\right\}.
%\end{aligned}
%\end{equation}

%where we define
%\begin{equation}
%\begin{aligned}
%\widetilde{\mathbf{V}}\triangleq \left|\tau_K\right|^2\beta\sum_{k=1}^{K}\mathbf{v}_{r,K,k}\mathbf{v}_{r,K,k}^H,
%\end{aligned}
%\end{equation}
%\begin{equation}
%\begin{aligned}
%\mathbf{b}\triangleq
%\sqrt{\beta\left(1+\gamma_{K}\right)}\tau_K^{\ast}\mathbf{v}_{r,K,K}.
%\end{aligned}
%\end{equation}

%Now, the complete procedure of finding the optimal auxiliary vectors $\bm{\gamma}^{\star}$ and $\bm{\tau}^{\star}$, the BS beamforming $\mathbf{w}_k^{\star}$, and RIS reflection matrices $\bm{\Phi}_1^{\star}, \bm{\Phi}_2^{\star}$ is straightforward.
Now, the complete procedure of finding the optimal auxiliary vectors $\bm{\gamma}^{\star}$ and $\bm{\tau}^{\star}$, the BS beamforming $\mathbf{w}_k^{\star}$, and RIS reflection matrices $\bm{\Phi}_1^{\star}, \bm{\Phi}_2^{\star}$ is straightforward. With appropriate initialization, we iteratively update $\bm{\gamma}$, $\bm{\tau}$, $\mathbf{w}_k$ and $\bm{\Phi}_1, \bm{\Phi}_2$ until convergence. For clarity, the proposed FP-based BS beamforming and RIS reflection design algorithm is summarized in Algorithm \ref{alg:BS}.

\begin{algorithm}[!t]\begin{small}
\caption{Joint Beamforming and RIS Reflection Design}
\label{alg:BS}
\begin{algorithmic}[1]
\REQUIRE $\mathbf{h}_k, \forall k$, $\mathbf{G}$, $\mathbf{g}_\mathrm{r}$, $\mathbf{g}_\mathrm{t}$, $b$, $\sigma^2$, $\sigma_0^2$, $\beta$, $P_\mathrm{T}$, $T_{max}$.
\ENSURE $\mathbf{w}_k^{\star}, \forall k$, $\bm{\Phi}_1^{\star}$, $\bm{\Phi}_2^{\star}$.
    \STATE {Initialize $\mathbf{w}_k, \forall k$, $\bm{\Phi}_1$, $\bm{\Phi}_2$.}
    \STATE{\textbf{Repeat}}
        \STATE{\hspace{0.4 cm}Update $\bm{\gamma}^{\star}$ by (\ref{eq:rrr}).}
        \STATE{\hspace{0.4 cm}Update $\bm{\tau}^{\star}$ by (\ref{eq:ttt}).}
        \STATE{\hspace{0.4 cm}Update each beamforming vector $\mathbf{w}_k^{\star}, \forall k$, by solving (\ref{eq:w}).}
        \STATE{\hspace{0.4 cm}Set $t=1$.}
        \STATE{\hspace{0.4 cm}\textbf{while} $t<T_{max}$ \& no convergence of $\bm{\Phi}_1$ do}
        %\STATE{\hspace{0.6 cm}\textbf{for} $m =1:M$ \textbf{do}}
        \STATE{\hspace{0.6 cm}Update $\bm{\phi}_{1}^{t+1}$ by (\ref{eq:psi1})/(\ref{eq:qpsi2}).}
        \STATE{\hspace{0.6 cm}$t:=t+1$.}
        %\STATE{\hspace{0.6 cm}\textbf{end for}}
        \STATE{\hspace{0.4 cm}\textbf{end while}}
        \STATE{\hspace{0.4 cm}Set $t=1$.}
        \STATE{\hspace{0.4 cm}\textbf{while} $t<T_{max}$ \& no convergence of $\bm{\Phi}_2$ do}
        %\STATE{\hspace{0.6 cm}\textbf{for} $m =1:M$ \textbf{do}}
        \STATE{\hspace{0.6 cm}Update $\bm{\phi}_{2}^{t+1}$ by (\ref{eq:psi2})/(\ref{eq:qpsi2}).}
        \STATE{\hspace{0.6 cm}$t:=t+1$.}
        %\STATE{\hspace{0.6 cm}\textbf{end for}}
        \STATE{\hspace{0.4 cm}\textbf{end while}}
    \STATE{\textbf{Until convergence}}
    \STATE{Return $\mathbf{w}_k^{\star}, \forall k$, $\bm{\Phi}_1^{\star}$, $\bm{\Phi}_2^{\star}$.}
    \end{algorithmic}\end{small}
\end{algorithm}

\subsection{Complexity Analysis}
In this subsection, we provide a brief computational complexity analysis for the proposed joint BS beamforming and RIS reflection design for the sum-rate maximization problem, i.e., Algorithm \ref{alg:BS}.

The overall computational complexity of proposed algorithm is mainly caused by the update of variables. In each iteration, obtaining the optimal solution of $\bm{\gamma}$ and $\bm{\tau}$ requires approximately $\mathcal{O}(K^2 M^2)$ and $\mathcal{O}(K(K+1) M^2)$ operations, respectively; updating the BS transmit beamforming $\mathbf{w}_k$ requires about $\mathcal{O}(M^2 N^{3})$ operations, which is mainly caused by the matrix inversion operation. Computing the RIS phase-shift matrices $\bm{\Phi}_1$ and $\bm{\Phi}_2$ has a complexity of approximately $\mathcal{O}(I_1M^{2})$ and $\mathcal{O}(I_2M^{2})$ with $I_1$ and $I_2$ being the number of iterations. Therefore, the total computational complexity of Algorithm \ref{alg:BS} can be approximated by $\mathcal{O}\left(I_R\left(M^2 N^{3}+I_1M^{2}+I_2M^{2}+K^2 M^2\right)\right)$, wherein $I_R$ represents the required number of iterations for the algorithm convergence.

\section{Simulation Results}
\label{sec:simulation results}
%\vspace{0.2 cm}

In this section, we provide simulation results to demonstrate the advancement of the proposed dual-functional RIS architecture and illustrate the effectiveness of our proposed algorithm.
%In the considered MU-MISO system,
%antenna spacing $d$ is $\lambda/2$ where $\lambda$ denotes the wavelength. T
The noise power at all receivers is $\sigma_k^2 = -80$dBm and the thermal noise power introduced by the amplifier is $\sigma_0^2 = -70$dBm \cite{Wu}.
The path-loss is modeled as $\text{PL}(d) = C_0\left(d_0/d\right)^\kappa$, where $C_0 = -30$dB, $d_0 = 1$m, $d$ is the link distance, and $\kappa$ is the path-loss exponent.
The BS-RIS channel is assumed to follow a small-scale Rician fading model with a Rician factor of 3dB and a path-loss exponent of 2.5, while the other channels only have non-line-of-sight (NLoS) components and a path-loss exponent of 3.
The channel $\mathbf{g}_\mathrm{t}$ from RIS-1 to horn antenna and the channel $\mathbf{g}_\mathrm{r}$ from horn antenna to RIS-2 adopt near-field model \cite{Cheng}, \cite{near-field}.
The BS is equipped with $N = 6$ antennas and serves $K = 4$ users.
The RIS is 50m away from the BS, while the three users ($k = 1,2,3$) are 2m away from the dual-functional RIS and the fourth user ($k=4$) is 20m away from the RIS.

\begin{figure}[!t]
\centering
\includegraphics[width = 3.52 in]{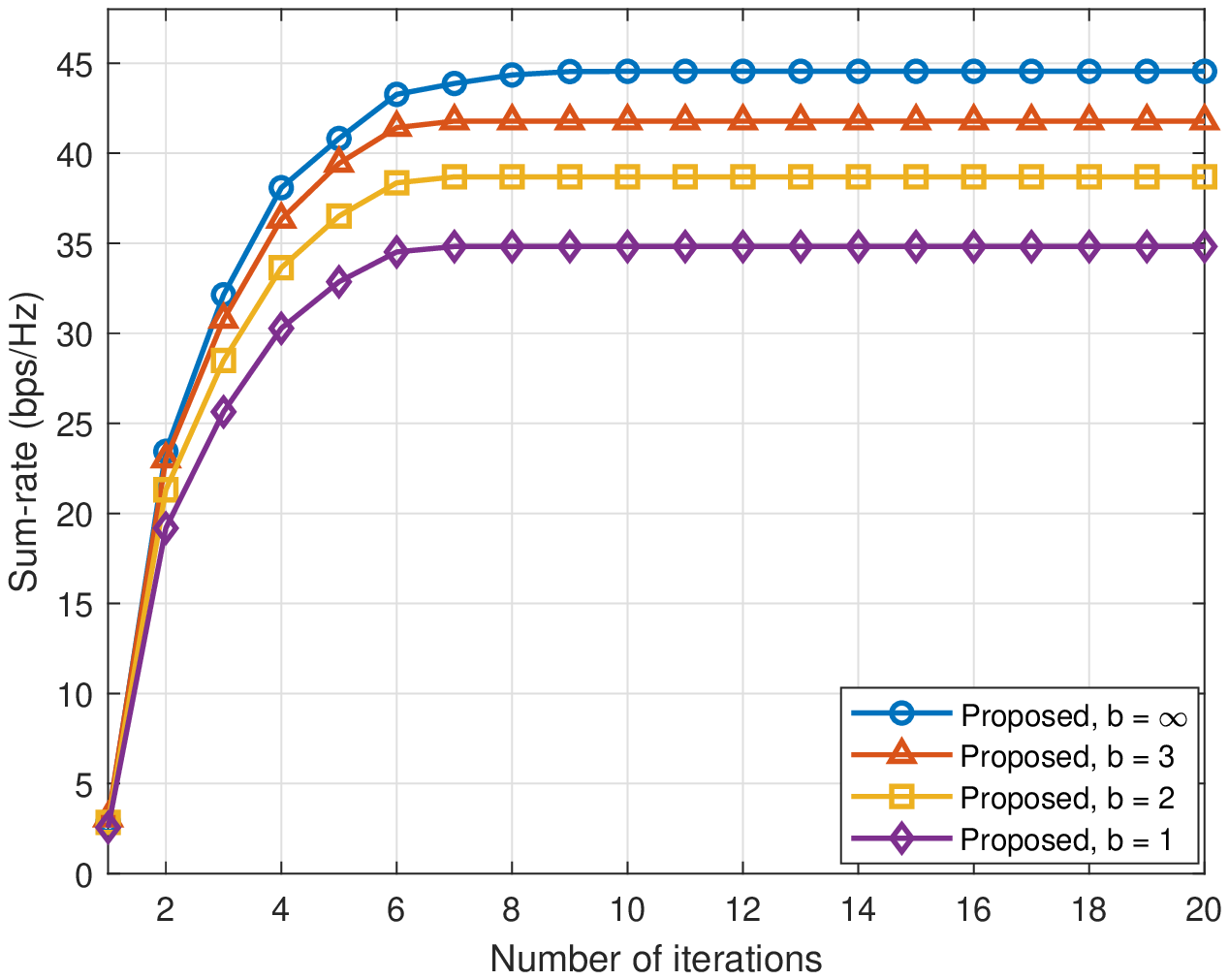}\vspace{0.1 cm}
\caption{Sum-rate versus the number of iterations ($P_\mathrm{T}$ = 40dBm, $M$ = 64, $\beta$ = 30dB).}\label{fig:conv}
\vspace{-0.0 cm}
\end{figure}

%\begin{figure}[!t]
%\centering
%\includegraphics[width = 3.0 in]{bit.eps}\vspace{0.1 cm}
%\caption{Sum-rate versus the resolution $b$ ($P_\mathrm{T}$ = 40dBm, $M$ = 64).}\label{fig:bit}
%\vspace{-0.2 cm}
%\end{figure}

\begin{figure}[!t]
\centering
\includegraphics[width = 3.52 in]{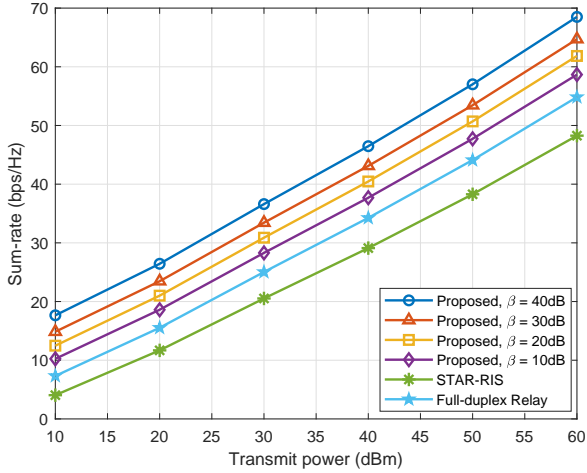}\vspace{0.1 cm}
\caption{Sum-rate versus the transmit power $P_\mathrm{T}$ for different amplifier gain $\beta$ ($M$ = 64, $b = \infty$. The BS power consumption of ``STAR-RIS'' is the same with the total power consumption of dual-functional RIS assisted system with $\beta$ = 40dB).}\label{fig:rate_beta}
\vspace{-0.0 cm}
\end{figure}

%
%\begin{figure}[!t]
%\centering
%\includegraphics[width = 3.52 in]{rvsp.eps}\vspace{0.1 cm}
%\caption{Sum-rate versus the transmit power $P_\mathrm{T}$ ($M$ = 64, $\beta$=30dB).}\label{fig:rate}
%\vspace{-0.0 cm}
%\end{figure}

We start with presenting the convergence of the proposed joint BS beamforming and RIS reflection design by plotting the sum-rate versus the number of iterations in Fig. \ref{fig:conv}. Simulation result illustrates that the proposed algorithm can converge fast and within 15 iterations.

Fig. \ref{fig:rate_beta} shows the sum-rate with the proposed dual-functional RIS architecture under different realistic values for the amplification gain $\beta$. For fair comparison, we also include the full-duplex relay equipped with $4$ antennas and STAR-RIS proposed in \cite{STAR} as the benchmark which consumes the same power as the proposed dual-functional RIS aided system. It is noted that the dual-functional RIS with a reasonable amplification gain can result in better performance compared with full-duplex relay since the large-scale RIS has significant beamforming gain. When the amplification gain $\beta$ is 40dB, only 30dBm transmit power is required to achieve the same performance, for example, sum-rate is around 37 bps/Hz, while 50dBm transmit power is need when STAR-RIS is deployed.

%Next, Fig. \ref{fig:rate} shows the achievable sum-rate versus transmit power $P_\mathrm{T}$ for the cases that RIS uses continuous and $b$ = 1, 2, 3-bit resolution phase shifters.
%It can be observed that the proposed dual-functional RIS architecture can achieve satisfactory performance
%%We can notice that the proposed dual-functional RIS can
%and can substantially enhance the communication quality compared with the conventional passive STAR-RIS scheme, which indicates the effectiveness of deploying dual-functional RIS.

\begin{figure}[!t]
\centering
\includegraphics[width = 3.52 in]{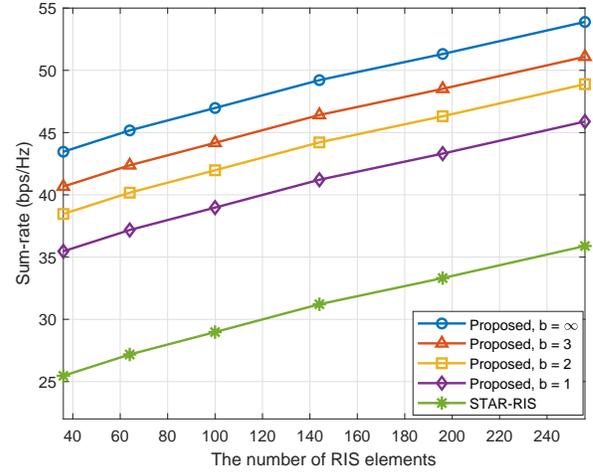}\vspace{0.1 cm}
\caption{Sum-rate versus the number of RIS elements $M$ ($P_\mathrm{T}$ = 40dBm, $\beta$ = 30dB).}\label{fig:ratenum}
\vspace{-0.0 cm}
\end{figure}

Fig. \ref{fig:ratenum} illustrates the sum-rate versus the number of RIS elements $M$. The same conclusion can be drawn that the proposed dual-functional RIS architecture can substantially enhance the communication quality compared with the conventional passive STAR-RIS scheme. As we can see, only 36 elements adopting 1-bit resolution phase shifters are needed for the proposed dual-functional RIS to achieve the same performance as passive STAR-RIS with 256 elements, which means the dual-functional RIS architecture can improve wireless communication performance without increasing the number of elements.
%It can be observed that the proposed dual-functional RIS architecture can substantially enhance the communication quality compared with the conventional passive STAR-RIS scheme, which indicates the effectiveness of deploying dual-functional RIS.
%When the RIS has continuous phase shifters, only 64 elements with amplification gain is $\beta =$30dB are needed to achieve the same performance as 100 elements with $\beta = 10$dB, which means the RIS-based MIMO relay architecture can improve wireless communication performance without increasing the number of elements.

\section{Conclusions}
\label{sec:conclusion}
\vspace{0.00 cm}

In this paper, we introduced a novel reflection and relay dual-functional RIS architecture to overcome the double fading effect and  enhance the coverage of downlink MU-MISO systems. Aiming at maximizing the achievable sum-rate, an iterative algorithm based on FP and MM was proposed to design the transmit beamforming and RIS reflection. Simulation results  demonstrated the superiority of the introduced dual-functional RIS  architecture and the effectiveness of the proposed algorithm.

%\vspace{-0.2 cm}

%\enlargethispage{-6.5cm}
\end{document}